\documentclass[runningheads]{llncs}
\usepackage[T1]{fontenc}
\usepackage{amssymb}

\usepackage{amsmath}
\usepackage{amsfonts}
\usepackage{bbm}
\usepackage{cleveref}

\usepackage{graphicx}

\newcommand{\hospitalMRI}{Klinikum Rechts der Isar (Munich)}
\newcommand{\hospitalLMU}{Klinikum der Universit\"at M\"unchen (Munich)}
\begin{document}

\title{Interpretable Vertebral Fracture Diagnosis}

\author{Paul Engstler\inst{1}\thanks{P. Engstler and M. Keicher - Both authors share first authorship.} \and
 Matthias Keicher\inst{1}\textsuperscript{\textasteriskcentered} \and
 David Schinz\inst{2} \and
 Kristina Mach\inst{1} \and
 Alexandra S. Gersing\inst{2,3} \and
 Sarah C. Foreman\inst{2} \and
 Sophia S. Goller\inst{3} \and
 Juergen Weissinger\inst{3} \and
 Jon Rischewski\inst{3} \and
 Anna-Sophia Dietrich\inst{2} \and
 Benedikt Wiestler\inst{2} \and
 Jan S. Kirschke\inst{2} \and
 Ashkan Khakzar\inst{1} \and
 Nassir Navab\inst{1,4}}

\authorrunning{P. Engstler et al.}
\institute{Technical University of Munich
\and
Klinikum Rechts der Isar (Technical University of Munich)
\and
Klinikum der Universität München (University of Munich)
\and
Johns Hopkins University}
\maketitle

\begin{abstract}
Do black-box neural network models learn clinically relevant features for fracture diagnosis? The answer not only establishes reliability quenches scientific curiosity but also leads to explainable and verbose findings that can assist the radiologists in the final and increase trust. This work identifies the concepts networks use for vertebral fracture diagnosis in CT images. This is achieved by associating concepts to neurons highly correlated with a specific diagnosis in the dataset. The concepts are either associated with neurons by radiologists pre-hoc or are visualized during a specific prediction and left for the user’s interpretation. We evaluate which concepts lead to correct diagnosis and which concepts lead to false positives. The proposed frameworks and analysis pave the way for reliable and explainable vertebral fracture diagnosis. The code is publicly available.\footnote{https://github.com/CAMP-eXplain-AI/Interpretable-Vertebral-Fracture-Diagnosis}

\end{abstract}

\section{Introduction}
Osteoporosis is regarded as one of the most relevant diseases of the elderly, with 22 million women and 5.5 million men affected in the EU alone \cite{cauley_public_2013,hernlund_osteoporosis_2013}. Early detection of incidental osteoporotic fractures in routinely-acquired computed tomography (CT) scans is important, as these often remain clinically silent for a long time \cite{haczynski_vertebral_2001}. Furthermore, osteoporotic fractures are an independent predictor of further fractures with an approx. 12-fold increased risk and are associated with an 8-fold increased mortality \cite{cauley2000risk,melton_iii_vertebral_1999}. The sequelae include major socioeconomic consequences and an individual reduction in quality of life \cite{bliuc_mortality_2009,jalava_association_2003,hallberg_health-related_2009,center_mortality_1999}. Despite the clinical significance, around 85\% of osteoporotic fractures are not adequately described in the radiological reports of routinely acquired CT scans, possibly as a result of a disproportionate increase in radiologists’ workload \cite{williams_under-reporting_2009,bartalena_prevalence_2009}.

Automatic detection of vertebral body fractures with deep learning models can remedy this and increase incidental findings. 
However, most of these methods are black-box models that do not give insights into the decision-making process. Revealing the inside of these models can allow for investigation of failure cases and, when addressed, increase robustness and trust in the system. 

Thus far, interpretable diagnosis is mostly investigated via feature attribution (saliency) approaches \cite{khakzar2021towards} such as class activation maps \cite{zhou2016learning}. These interpretations reveal where important features for the prediction are located. Although being a valuable tool for running a sanity check on the network inference mechanism, feature attribution does not disclose further information regarding prediction. Moreover, only knowing about the location of important features is not useful information for fracture diagnosis as it is easy to see where the fracture is located, and it is of interest to know “what” features are important.

To this end, inspired by the network dissection \cite{bau2017network} approach and its derivations in chest radiography \cite{khakzar2021towards} and mammography \cite{wu2018deepminer} applications, we propose two scenarios for analyzing the internal units of the neural network and their associated clinical concepts. The scenarios are dataset-wise and single inference. In the dataset-wise scenario, we compute the output of the last convolutional layer for all input data and identify neurons highly correlated with the output value associated with fractures. Subsequently, we ask the clinicians to identify the concepts associated with highly correlated activations by inspecting the inputs that activate those neurons the highest. The dataset-wise scenario provides an overall understanding of what concepts the network has learned and whether they are aligned with what clinicians use. In the single-inference scenario, the highly activated convolutional neurons for a single input are identified. Then their associated concepts are visualized to the user by showing the top images that activate each neuron. In this scenario, the user can get a conceptual understanding of the decision-making mechanism of the model. We perform the analysis for both scenarios on the open-source VerSe \cite{sekuboyina2021verse} dataset and a larger private dataset procured in our hospital. The concept-based interpretations are the building block toward a broader objective of explainable diagnosis and generating radiology reports. The objective of this work is to investigate what features the network uses for fracture diagnosis, whether they overlap with clinical knowledge, and how they can be used for more verbose and explainable fracture diagnosis.

\subsection{Related Work}
\subsubsection{Vertebral Fracture Detection}
Many works have been proposed for automatic vertebral fracture detection in recent years. Most of these approaches use Convolutional Neural Networks (CNN) on Computer Tomography (CT) spine images. Notable exceptions are \cite{valentinitsch2019random_forest} using 3D radiomics extracted from CT images in a Random Forest and methods \cite{murata2020radiograph,chen2021radiograph} detecting fractures on radiographs. 

CNN-based methods can be categorized into 2D and 3D convolutions. 2D methods usually rely on a feature aggregation with Recurrent Neural Networks to model inter-slice dependencies \cite{bar2017_2D-CNN_RNN,tomita2018_2D-CNN_LSTM}. Husseini et al. \cite{husseini2020grading} reformat the image to use the most informative mid-sagittal slice of each vertebra and, in addition to fracture detection, grade fractures using an ordinal regression loss for representation learning. Pisov et al. \cite{pisov2020keypoints} also reformat the 3D volume to retrieve a spine-centered 2D image and detect key points for measuring the compression of each vertebra, detecting and grading fractures. 

Detecting fractures on a voxel-level and then post-processing, Nicolas et al. \cite{nicolaes2019detection} for the first time used 3D convolutions for the detection of vertebral fractures. More recent works using 3D convolutions include modeling the dependency between the 3D volumes of each vertebra with a sequence-to-sequence model \cite{chettrit20203sequence3d}  and detecting osteoporotic fractures on a patient-level \cite{yilmaz2021automated}.
Related to the task of fracture detection and grading, recently Li et al. \cite{li2021malign_resnet50}, and Feng et al. \cite{feng2021malign_gradcam} explored the distinction between benign and malign vertebral fractures.

\subsubsection{Interpretability}
 of models is narrowly explored in the domain of vertebral fracture diagnosis and \cite{yilmaz20203dunet_feat_attr} interprets the models by feature attribution (saliency) approaches to identify which regions in the input contributed to the prediction. In fact, in most medical image analysis applications, feature attribution is the dominant approach \cite{khakzar2021towards}. However, attribution methods are limited in the information they can disclose regarding the decision-making mechanism of the model. Moreover, the feature attribution problem remains largely unsolved, and although there are many attribution approaches (CAM~\cite{zhou2016learning}, LRP~\cite{Montavon2017}, DeepSHAP~\cite{Lundberg2017}, IBA~\cite{Schulz2020Restricting,zhang2021fine,khakzar2021explaining}…), the methods disagree with the identified important features \cite{khakzar2022explanations,zhang2021fine,khakzar2020rethinking}. This disagreement problem is a caveat for domain experts while utilizing these attribution methods. Thus there is a need for interpretation approaches that are reliable and reveal more information than “which region is important.” An inspiring approach, Network Dissection \cite{bau2017network}, identifies the concepts encoded by internal units (neurons) of the network. Motivated by this approach, Wu et al. \cite{wu2018deepminer} identify concepts the network encodes for diagnosis on mammography images, and Khakzar et al. \cite{khakzar2021towards} perform dissection on chest x-ray models and investigate research questions such as what clinical concepts do networks pick up when trained on COVID-19 severity scores. Methodologically, our work differs from \cite{khakzar2021towards,bau2017network} in that we do not use an annotation dataset and instead identify the highly correlated neurons with the output under investigation. We investigate a different medical domain and explore different research questions such as what features contribute to true positives and what features to false positives.

\section{Methodology}

\subsection{Vertebral Fracture Detection}
We model the vertebral fracture detection task as a binary classification problem, where the positive class indicates a fracture. The network function is defined as $f_{\Theta}(x) : \mathbb{R}^{H \times W \times D} \to \mathbb{R}$. The predicted probability is $\hat{y} = sigmoid(f_{\Theta}(x))$. We use a 3D U-Net \cite{cciccek20163d} for the vertebral fracture classification task, replacing its upsampling path with a classification head. 

\subsection{Semantic Concept Extraction (Correlation)}
\label{ssec:correlation}
In neural networks, each neuron is activated by a specific input pattern. The corresponding pattern of each neuron can be equivalently deemed as its associated \emph{concept}. In convolutional neuron networks each neuron can be considered either as an activation map or an activation unit within the map. As the activation units within an activation map all represent the same function (only for different spatial locations), they represent the same concept \cite{bau2017network}.
We denote the output activations of the final convolutional layer of the network by the tensor $A \in \mathbb{R}^{H^{'}\times W^{'}\times K}$ where $K$ represents the number of channels in that layer.
After computing the distribution of individual unit activations
$a_k$, we determine the top quantile level $\mathcal{T}_k$ for each unit $k$ such that $P(a_k > \mathcal{T}_k) = 0.005$ \cite{bau2017network}. We then derive the binary segmentation mask $M_k(\pmb{x}) := A_k(\pmb{x}) > \mathcal{T}_k$ and denote the set of enabled units for an input $\pmb{x}$ as ${E_x} := \{k \mid \sum M_k(\pmb{x}) > 0\}$.

\subsubsection*{Positive Prediction Correlation}
Some units might capture concepts that are highly useful to determine whether a sample is fractured, establishing a stronger correlation with a true positive prediction than other units. To find these units, we compute:

\begin{equation}
    c_k := \frac{\sum_{x \in P} \pmb{1}_{E_x}(k)}{\lvert P \lvert}
\end{equation}

\noindent
where $P$ is the set of positive samples and $\pmb{1}$ is the indicator function. With $c_{k_1} > c_{k_2} > ...$, $k_1$ is the unit most strongly correlated with a true positive prediction, followed by $k_2$.

\subsection{Visualization of Highly Correlating Concepts at Inference}
\label{ssec:inference-importance}
Due to the variability of observed defects in fractured vertebrae, different concepts are relevant during the inference of a sample. We compute the relevance of a unit $k$ during inference of input $\pmb{x}$ as follows:

\begin{equation}
    r_k := \sum M_k(\pmb{x}) \odot A_k(\pmb{x})
\end{equation}

\noindent
For units $k_1$, $k_2$ with $r_{k_1} > r_{k_2}$, $k_1$ is more relevant for the inference of $\pmb{x}$ than $k_2$. Now, when visualizing highly correlated concepts for a sample $\pmb{x}$, we compute the inference relevance of each detector unit and display the activation maps $A_{k_1}(x)$, $A_{k_2}(x)$, ... with $r_{k_1} > r_{k_2} > ...$, showing the corresponding responses for the input sample $\pmb{x}$.
\section{Experimental Setup}

\subsubsection*{Data Preparation}
The network is trained on the VerSe dataset 
\cite{sekuboyina2021verse}
as well as an in-house dataset acquired at {\hospitalMRI} and {\hospitalLMU}.
The latter includes 465 patients with a median age of $\sim 69 (\pm12)$ years, containing a heterogeneous collection of field of views, scanner settings, and healthy and fractured vertebra, including metallic implants and foreign materials.
This combined dataset contains CT scans of patients with healthy and fractured vertebrae of osteoporotic or malignant nature from a heterogeneous collection of CT scanners.
To address the inherent class imbalance in the data, 
negative samples are undersampled and positive (fractured) samples are oversampled in training to achieve a perfect class balance each epoch. As osteoporotic and malignant fractures rarely occur in cervical vertebrae (C1-C7), they are excluded from the dataset. We extract $96\times96\times96$ sized 3D patches for each vertebrae with a 1mm resolution. These patches are centered on the vertebral body and oriented along the spine by aligning the vertical axis with a spline constructed with the vertebral centroids provided by the dataset similar to \cite{husseini2020grading}. The intensity values of the resulting crops are cropped to a Hounsfield Unit range of $[-1000, 1000]$ and then scaled to $[0, 1]$. During training, intensity (Gaussian noise, smoothing, and contrast) and heavy spatial data augmentations (similarity transformation and elastic deformation) are applied. For these tasks, NiBabel 3.2.1 and MONAI 0.8.0 are used. 

\subsubsection*{Implementation Details}
 The 3D U-Net is implemented in PyTorch Lightning 1.5.10 on top of PyTorch 1.10.2, and trained using the Adam \cite{kingma2014adam} optimizer (learning rate 0.001) without weight decay. Training is concluded if the validation F1 score has not improved for 50 epochs. Dropout with probability 0.3 is applied.
\section{Results and Discussion}
In the following, we first evaluate the performance of our vertebral fracture detection neural network before dissecting it into its individual detector units. We then validate detector units highly correlated with a true positive prediction by showing that they represent clinically meaningful concepts. Lastly, we present a system to display the units most relevant to a single inference.

\subsubsection*{Vertebral Fracture Detection} We consider the threshold-based evaluation metrics F1-score and accuracy. To remove the dependence on a manually chosen threshold whose optimum might vary between trained networks, the area under curve (AUC) and average precision (AP) metrics are also evaluated. We report the mean and standard deviation of these metrics from five separate training trials for each model.

\begin{table}
\begin{center}
{\def\arraystretch{1.25}\tabcolsep=3pt
\begin{tabular}{c c c c c c}
Training & Testing & F1 (\%) & Acc. (\%) & AUC (\%) & AP (\%)\\
\hline
VerSe & VerSe & $71.2\pm10.8$ & $78.2\pm12.0$ & $84.5\pm9.1$ & $76.4\pm14.5$ \\
VerSe, in-house & VerSe & $86.1\pm2.6$ & $\pmb{90.9\pm1.6}$ & $\pmb{96.2\pm0.9}$ & $94.1\pm1.6$ \\
VerSe, in-house & VerSe, in-house & $\pmb{88.0\pm0.7}$ & $88.0\pm0.4$ & $94.7\pm0.5$ & $\pmb{95.0\pm0.4}$\\
\end{tabular}}
\end{center}
\caption{Performance of the trained neural networks on the test holdout of the smaller VerSe dataset as well as the combined dataset, comprised of VerSe and non-public data acquired from {\hospitalMRI} and {\hospitalLMU}. In total, the VerSe dataset contains 3,920 non-cervical vertebrae (254 of which are fractured), whereas the combined dataset comprises 10,675 T1-L5 vertebrae (1,246 fractured).}
\end{table}

For networks trained on the smaller VerSe dataset, we observe performance akin to "naive" two-dimensional vertebral fracture detection approaches on the same dataset \cite{husseini2020grading}, and a high dependence on a beneficial random seed. These networks, however, do not yield detector units that exhibit any discerning patterns.
This is achieved by training a network with the larger dataset, combining VerSe and in-house data collected at {\hospitalMRI} and {\hospitalLMU}, that is reliably superior in performance. Its detector units exhibit a variety of patterns that are investigated in the subsequent sections.

\begin{table}[h!]
\begin{center}
\tiny
{\def\arraystretch{2}\tabcolsep=5pt
\begin{tabular}{@{\hspace{0\tabcolsep}} l @{\hspace{0.5\tabcolsep}} l p{3cm}}
\hline
Rank & Sample Activations & Clinical Explanation\\
\hline
1 & \raisebox{-.85\height}{\includegraphics[width=0.65\textwidth]{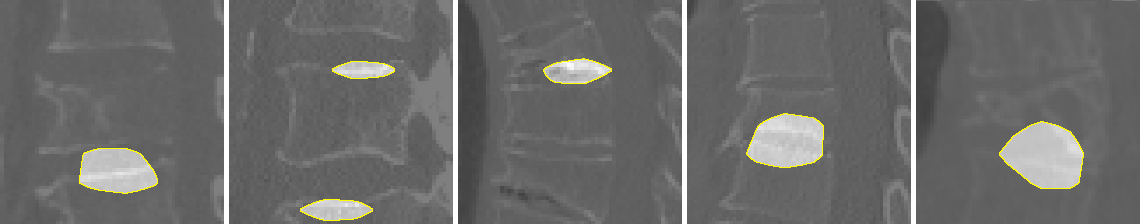}} & Abnormal endplate and intervertebral disc shapes \\

2 & \raisebox{-.85\height}{\includegraphics[width=0.65\textwidth]{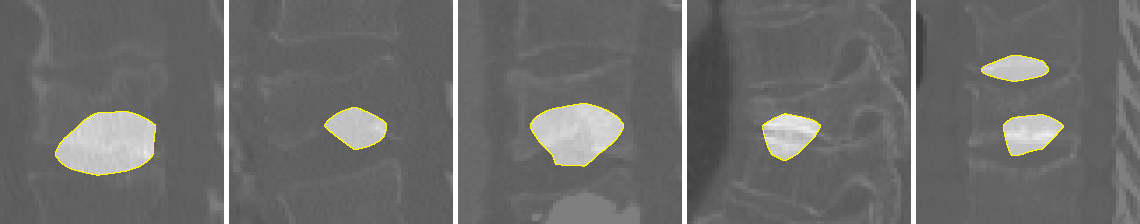}} & Primarily defects of the inferior endplate, associated with severe fractures\\ 

5 & \raisebox{-.85\height}{\includegraphics[width=0.65\textwidth]{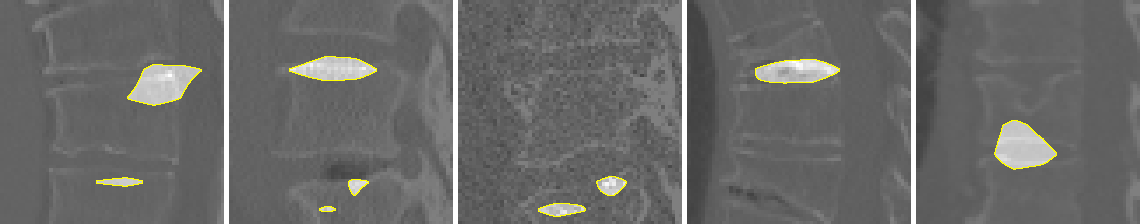}} & Abnormal endplate shapes with partial observation of adjacent inferior vertebrae \\

7 & \raisebox{-.85\height}{\includegraphics[width=0.65\textwidth]{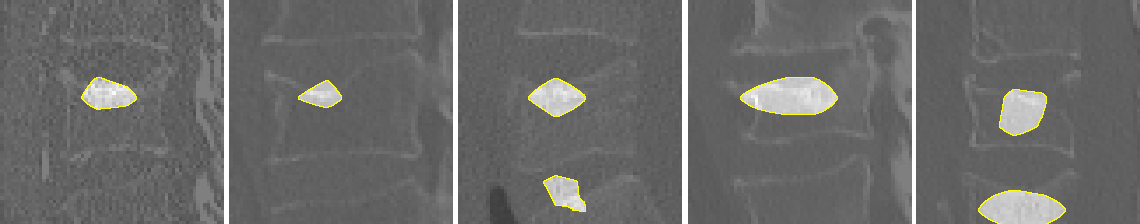}} & Central defect of the superior endplate, commonly observed in compression fractures, with partial observation of adjacent inferior vertebrae \\

8 & \raisebox{-.85\height}{\includegraphics[width=0.65\textwidth]{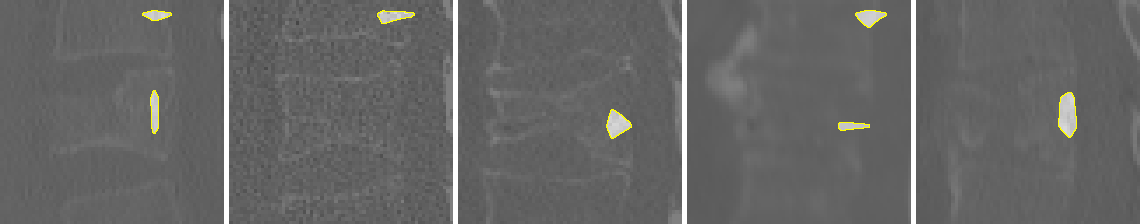}} & Observation of the spongiosa in the primary vertebrae as well as the adjacent superior one \\

9 & \raisebox{-.85\height}{\includegraphics[width=0.65\textwidth]{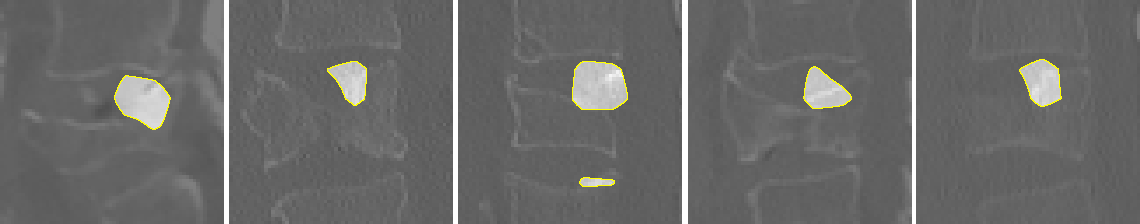}} & Injury to the middle column of the vertebral bodies, associated with clinically significant myelon compression and consecutive paresis \\

10 & \raisebox{-.85\height}{\includegraphics[width=0.65\textwidth]{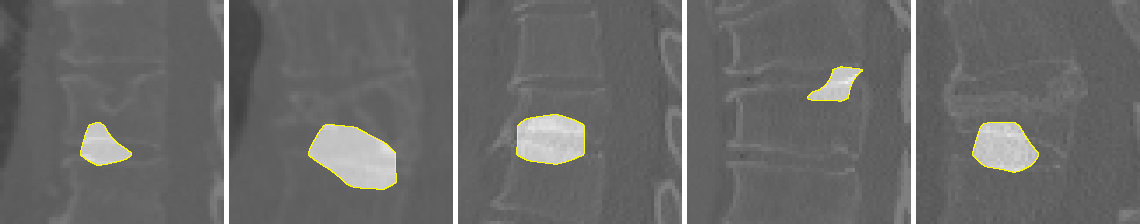}} & Abnormal endplate and intervertebral disc shapes \\

\end{tabular}}
\end{center}
\caption{Visualization of the detector units most strongly correlated with a true positive prediction along with an interpretation of their activations by clinical experts. All displayed samples are fractured and represented by a slice with high activation after thresholding.}
\label{tab:top-units-overview}
\end{table}

\subsection{Clinical Meaningfulness of Extracted Semantic Concepts}
Given the network trained on the larger dataset, we extract its semantic concepts with Network Dissection \cite{bau2017network}, which we extended to the three-dimensional space. To reduce the 512 detector units of the 3D U-Net to a tractable number, we determine the top ten units highly correlated with a true positive prediction as detailed in \Cref{ssec:correlation}. For these units, we exported a single-slice collage of 25 strongly activating fractured samples serving as an overview of the units' activations. For the five samples that activated the unit most strongly, all two-dimensional slices as well as three-dimensional NIfTI files are exported, allowing for a detailed inspection.

Based on these exports, we consulted two clinical experts with a combined experience of 22 years in spine imaging about the clinical meaningfulness of these detector units. Omitting three units where no immediate association was possible, we show the detector units identified by their correlation rank with their corresponding clinical explanation in \Cref{tab:top-units-overview}. The provided samples show a diverse collection of detector unit activations, with each unit exhibiting consistent patterns across multiple samples. We also observe that these units' main focus is the primary vertebra, even if there is some activation in the surroundings. It is noteworthy that the patterns align with the bone anatomy and present themselves in clinically significant locations. As severe fractures are associated with changes in the superior and inferior vertebral endplates, we find the majority of activations in these regions. Although multiple detector units target these areas, they focus on different locations and exhibit varying sizes of regions of interest, with some integrating further information from the intervertebral discs as well as the adjacent vertebra. These insights are clinically meaningful to detect moderate and severe vertebral deformations (Genant grade 1 or higher \cite{genant1993vertebral}), and thus show that our network learned concepts that have a clinical correspondence.

For the omitted cases, we observed either no statistically significant activations, i.e. $M_{k}(\pmb{x}) = \pmb{0}$, or sporadic activations that do not present any clear patterns, even though they are highly correlated with a true positive prediction. Overall, such detector units represent a minority and can therefore be disregarded in light of those that exhibit tangible patterns. 

\subsection{Single-Inference Concept Visualization}
Having shown that the network learns clinically relevant concepts, we have validated its ability to make use of conducive features. We further seek to illuminate the black box decision-making process of the network by providing the user with a visual explanation for a single inference. To this end, we propose a system that visualizes the concepts considered most important by the network during inference.

Using the method described in \Cref{ssec:inference-importance} to identify the units representing the most relevant concepts, we retrieve their respective top activating images from our combined dataset. We then display two visualizations for each unit: (i) the activations of those units for the input sample, and (ii) the activations for their corresponding top images. This provides the user with a detector unit's particular response for the given input sample as well as a larger context to understand its general concept. For both visualizations, a single slice with high activation (after thresholding) is shown. An example of (i) is given with \Cref{tab:inference-visualization}, which gives evidence of the network corroborating its prediction with a diverse set of concepts. These concepts illustrate the network accurately identifying relevant indications for the wedge-shaped deformity and incorporating information from an adjacent vertebra.

\begin{table}
\begin{center}
{\def\arraystretch{1.5}\tabcolsep=2pt
\begin{tabular}{@{\hspace{0\tabcolsep}} c @{\hspace{5\tabcolsep}} ccccccc}
\raisebox{-.85\height}{\includegraphics[height=1.5cm]{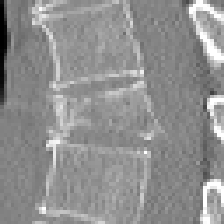}} & \raisebox{-.85\height}{\includegraphics[height=1.5cm]{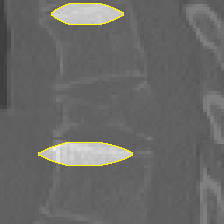}} & \raisebox{-.85\height}{\includegraphics[height=1.5cm]{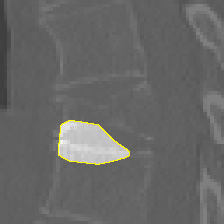}} & \raisebox{-.85\height}{\includegraphics[height=1.5cm]{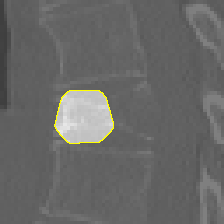}} & \raisebox{-.85\height}{\includegraphics[height=1.5cm]{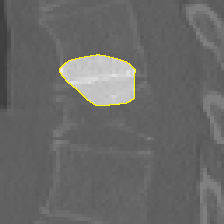}} &
\raisebox{-.85\height}{\includegraphics[height=1.5cm]{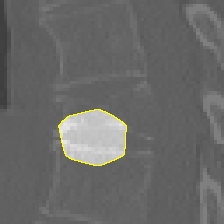}} &
& \raisebox{-.85\height}{\includegraphics[height=1.5cm]{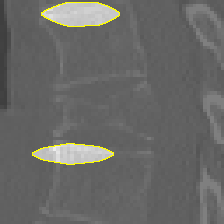}} \\
Unit & 22 & 111 & 301 & 122 & 277 &  & 197 \\
Relevance & 1 & 2 & 3 & 4 & 5 & \dots  & 10 \\
Corr. Rank & 280 & 33 & 149 & 29 & 162 & & 466 \\
\end{tabular}}
\end{center}
\caption{Visualization of the most relevant detector units during class prediction of the sample shown on the left, which the network correctly predicted as fractured. Each detector unit is represented by a single slice activation for that particular sample. We also show its ranking in units highly correlated with a true positive prediction. We observe that the network uses concepts associated with wedge-shaped deformity and incorporates information from an adjacent vertebra}
\label{tab:inference-visualization}
\end{table}

This system enables users to comprehend the network's decision making, increasing trust in the system and allowing them to identify failure cases more easily.
Furthermore, this approach does not require any prior concept matching by experts, as the user is able to interpret the general concept of a detector unit and make informed judgements about its importance for a particular sample.
\section{Conclusion}
We show that a 3D U-Net learns a diverse set of concepts to tackle the task of detecting vertebral fractures. To gauge their meaningfulness, we first proposed a method to identify units highly correlated with a fracture detection. Then, we showed the overlap of these units with clinical concepts as validated by experts. Finally, we introduced a system to visually explain a single inference by showing the concepts most relevant for the classification of the sample, giving users insight into the network's decision making process. Further extensions of this system are conceivable, such as pre-filling a radiology report based on activations in a group of semantically similar detector units.

\section*{Acknowledgements}
\label{sec:acknowledgements}
\noindent
The authors acknowledge the financial support by the Federal Ministry of Education and Research of Germany (BMBF) under project DIVA (FKZ 13GW0469C). Ashkan Khakzar was partially supported by the Munich Center for Machine Learning (MCML) with funding from the BMBF under the project 01IS18036B. Kristina Mach was partially supported by the Linde \& Munich Data Science Institute, Technical University of Munich PhD Fellowship.
\bibliographystyle{splncs04}
\bibliography{bibliography}

\begin{thebibliography}{10}
\providecommand{\url}[1]{\texttt{#1}}
\providecommand{\urlprefix}{URL }
\providecommand{\doi}[1]{https://doi.org/#1}

\bibitem{bar2017_2D-CNN_RNN}
Bar, A., Wolf, L., Amitai, O.B., Toledano, E., Elnekave, E.: Compression
  fractures detection on ct. In: Medical imaging 2017: computer-aided
  diagnosis. vol. 10134, p. 1013440. International Society for Optics and
  Photonics (2017)

\bibitem{bartalena_prevalence_2009}
Bartalena, T., Giannelli, G., Rinaldi, M.F., Rimondi, E., Rinaldi, G.,
  Sverzellati, N., Gavelli, G.: Prevalence of thoracolumbar vertebral fractures
  on multidetector {CT}. European Journal of Radiology  \textbf{69}(3),
  555--559 (Mar 2009)

\bibitem{bau2017network}
Bau, D., Zhou, B., Khosla, A., Oliva, A., Torralba, A.: Network dissection:
  Quantifying interpretability of deep visual representations. In: Proceedings
  of the IEEE conference on computer vision and pattern recognition. pp.
  6541--6549 (2017)

\bibitem{bliuc_mortality_2009}
Bliuc, D.: Mortality {Risk} {Associated} {With} {Low}-{Trauma} {Osteoporotic}
  {Fracture} and {Subsequent} {Fracture} in {Men} and {Women}. JAMA
  \textbf{301}(5), ~513 (Feb 2009). \doi{10.1001/jama.2009.50}

\bibitem{cauley_public_2013}
Cauley, J.A.: Public {Health} {Impact} of {Osteoporosis}. The Journals of
  Gerontology Series A: Biological Sciences and Medical Sciences
  \textbf{68}(10),  1243--1251 (Oct 2013)

\bibitem{cauley2000risk}
Cauley, J., Thompson, D., Ensrud, K., Scott, J., Black, D.: Risk of mortality
  following clinical fractures. Osteoporosis international  \textbf{11}(7),
  556--561 (2000)

\bibitem{center_mortality_1999}
Center, J.R., Nguyen, T.V., Schneider, D., Sambrook, P.N., Eisman, J.A.:
  Mortality after all major types of osteoporotic fracture in men and women: an
  observational study. The Lancet  \textbf{353}(9156),  878--882 (Mar 1999).
  \doi{10.1016/S0140-6736(98)09075-8}

\bibitem{chen2021radiograph}
Chen, H.Y., Hsu, B.W.Y., Yin, Y.K., Lin, F.H., Yang, T.H., Yang, R.S., Lee,
  C.K., Tseng, V.S.: Application of deep learning algorithm to detect and
  visualize vertebral fractures on plain frontal radiographs. Plos one
  \textbf{16}(1),  e0245992 (2021)

\bibitem{chettrit20203sequence3d}
Chettrit, D., Meir, T., Lebel, H., Orlovsky, M., Gordon, R., Akselrod-Ballin,
  A., Bar, A.: 3d convolutional sequence to sequence model for vertebral
  compression fractures identification in ct. In: International Conference on
  Medical Image Computing and Computer-Assisted Intervention. pp. 743--752.
  Springer (2020)

\bibitem{cciccek20163d}
{\c{C}}i{\c{c}}ek, {\"O}., Abdulkadir, A., Lienkamp, S.S., Brox, T.,
  Ronneberger, O.: 3d u-net: learning dense volumetric segmentation from sparse
  annotation. In: MICCAI. pp. 424--432. Springer (2016)

\bibitem{feng2021malign_gradcam}
Feng, S., Liu, B., Zhang, Y., Zhang, X., Li, Y.: Two-stream compare and
  contrast network for vertebral compression fracture diagnosis. IEEE
  Transactions on Medical Imaging  \textbf{40}(9),  2496--2506 (2021)

\bibitem{genant1993vertebral}
Genant, H.K., Wu, C.Y., Van~Kuijk, C., Nevitt, M.C.: Vertebral fracture
  assessment using a semiquantitative technique. Journal of bone and mineral
  research  \textbf{8}(9),  1137--1148 (1993)

\bibitem{haczynski_vertebral_2001}
Haczynski, J., Jakimiuk, A.: Vertebral fractures: a hidden problem of
  osteoporosis. Medical Science Monitor: International Medical Journal of
  Experimental and Clinical Research  \textbf{7}(5),  1108--1117 (Oct 2001)

\bibitem{hallberg_health-related_2009}
Hallberg, I., Bachrach-Lindström, M., Hammerby, S., Toss, G., Ek, A.C.:
  Health-related quality of life after vertebral or hip fracture: a seven-year
  follow-up study. BMC Musculoskeletal Disorders  \textbf{10}(1), ~135 (Dec
  2009). \doi{10.1186/1471-2474-10-135}

\bibitem{hernlund_osteoporosis_2013}
Hernlund, E., Svedbom, A., Ivergård, M., Compston, J., Cooper, C., Stenmark,
  J., McCloskey, E.V., Jönsson, B., Kanis, J.A.: Osteoporosis in the
  {European} {Union}: medical management, epidemiology and economic burden: {A}
  report prepared in collaboration with the {International} {Osteoporosis}
  {Foundation} ({IOF}) and the {European} {Federation} of {Pharmaceutical}
  {Industry} {Associations} ({EFPIA}). Archives of Osteoporosis
  \textbf{8}(1-2), ~136 (Dec 2013). \doi{10.1007/s11657-013-0136-1}

\bibitem{husseini2020grading}
Husseini, M., Sekuboyina, A., Loeffler, M., Navarro, F., Menze, B.H., Kirschke,
  J.S.: Grading loss: a fracture grade-based metric loss for vertebral fracture
  detection. In: MICCAI. Springer (2020)

\bibitem{jalava_association_2003}
Jalava, T., Sarna, S., Pylkkänen, L., Mawer, B., Kanis, J.A., Selby, P.,
  Davies, M., Adams, J., Francis, R.M., Robinson, J., McCloskey, E.:
  Association {Between} {Vertebral} {Fracture} and {Increased} {Mortality} in
  {Osteoporotic} {Patients}. Journal of Bone and Mineral Research
  \textbf{18}(7),  1254--1260 (Jul 2003). \doi{10.1359/jbmr.2003.18.7.1254}

\bibitem{khakzar2020rethinking}
Khakzar, A., Baselizadeh, S., Navab, N.: Rethinking positive aggregation and
  propagation of gradients in gradient-based saliency methods. arXiv preprint
  arXiv:2012.00362  (2020)

\bibitem{khakzar2022explanations}
Khakzar, A., Khorsandi, P., Nobahari, R., Navab, N.: Do explanations explain?
  model knows best. arXiv preprint arXiv:2203.02269  (2022)

\bibitem{khakzar2021towards}
Khakzar, A., Musatian, S., Buchberger, J., Valeriano~Quiroz, I., Pinger, N.,
  Baselizadeh, S., Kim, S.T., Navab, N.: Towards semantic interpretation of
  thoracic disease and covid-19 diagnosis models. In: MICCAI. Springer (2021)

\bibitem{khakzar2021explaining}
Khakzar, A., Zhang, Y., Mansour, W., Cai, Y., Li, Y., Zhang, Y., Kim, S.T.,
  Navab, N.: Explaining covid-19 and thoracic pathology model predictions by
  identifying informative input features. In: International Conference on
  Medical Image Computing and Computer-Assisted Intervention. pp. 391--401.
  Springer (2021)

\bibitem{kingma2014adam}
Kingma, D.P., Ba, J.: Adam: A method for stochastic optimization. arXiv
  preprint arXiv:1412.6980  (2014)

\bibitem{li2021malign_resnet50}
Li, Y., Zhang, Y., Zhang, E., Chen, Y., Wang, Q., Liu, K., Yu, H.J., Yuan, H.,
  Lang, N., Su, M.Y.: Differential diagnosis of benign and malignant vertebral
  fracture on ct using deep learning. European Radiology  \textbf{31}(12),
  9612--9619 (2021)

\bibitem{Lundberg2017}
Lundberg, S.M., Lee, S.I.: {A unified approach to interpreting model
  predictions}. In: Advances in Neural Information Processing Systems (2017)

\bibitem{melton_iii_vertebral_1999}
Melton~III, L.J., Atkinson, E.J., Cooper, C., O'Fallon, W.M., Riggs, B.L.:
  Vertebral {Fractures} {Predict} {Subsequent} {Fractures}. Osteoporosis
  International  \textbf{10}(3),  214--221 (Sep 1999).
  \doi{10.1007/s001980050218}

\bibitem{Montavon2017}
Montavon, G., Lapuschkin, S., Binder, A., Samek, W., M{\"{u}}ller, K.R.:
  {Explaining nonlinear classification decisions with deep Taylor
  decomposition}. Pattern Recognition  (2017).
  \doi{10.1016/j.patcog.2016.11.008}

\bibitem{murata2020radiograph}
Murata, K., Endo, K., Aihara, T., Suzuki, H., Sawaji, Y., Matsuoka, Y.,
  Nishimura, H., Takamatsu, T., Konishi, T., Maekawa, A., et~al.: Artificial
  intelligence for the detection of vertebral fractures on plain spinal
  radiography. Scientific Reports  \textbf{10}(1), ~1--8 (2020)

\bibitem{nicolaes2019detection}
Nicolaes, J., Raeymaeckers, S., Robben, D., Wilms, G., Vandermeulen, D.,
  Libanati, C., Debois, M.: Detection of vertebral fractures in ct using 3d
  convolutional neural networks. In: International Workshop and Challenge on
  Computational Methods and Clinical Applications for Spine Imaging. pp. 3--14.
  Springer (2019)

\bibitem{pisov2020keypoints}
Pisov, M., Kondratenko, V., Zakharov, A., Petraikin, A., Gombolevskiy, V.,
  Morozov, S., Belyaev, M.: Keypoints localization for joint vertebra detection
  and fracture severity quantification. In: MICCAI. pp. 723--732. Springer
  (2020)

\bibitem{Schulz2020Restricting}
Schulz, K., Sixt, L., Tombari, F., Landgraf, T.: Restricting the flow:
  Information bottlenecks for attribution. In: International Conference on
  Learning Representations (2020),
  \url{https://openreview.net/forum?id=S1xWh1rYwB}

\bibitem{sekuboyina2021verse}
Sekuboyina, A., Husseini, M.E., Bayat, A., L{\"o}ffler, M., Liebl, H., Li, H.,
  Tetteh, G., Kuka{\v{c}}ka, J., Payer, C., {\v{S}}tern, D., et~al.: Verse: a
  vertebrae labelling and segmentation benchmark for multi-detector ct images.
  Medical image analysis  (2021)

\bibitem{tomita2018_2D-CNN_LSTM}
Tomita, N., Cheung, Y.Y., Hassanpour, S.: Deep neural networks for automatic
  detection of osteoporotic vertebral fractures on ct scans. Computers in
  biology and medicine  \textbf{98},  8--15 (2018)

\bibitem{valentinitsch2019random_forest}
Valentinitsch, A., Trebeschi, S., Kaesmacher, J., Lorenz, C., L{\"o}ffler, M.,
  Zimmer, C., Baum, T., Kirschke, J.: Opportunistic osteoporosis screening in
  multi-detector ct images via local classification of textures. Osteoporosis
  international  \textbf{30}(6),  1275--1285 (2019)

\bibitem{williams_under-reporting_2009}
Williams, A.L., Al-Busaidi, A., Sparrow, P.J., Adams, J.E., Whitehouse, R.W.:
  Under-reporting of osteoporotic vertebral fractures on computed tomography.
  European Journal of Radiology  \textbf{69}(1),  179--183 (Jan 2009)

\bibitem{wu2018deepminer}
Wu, J., Zhou, B., Peck, D., Hsieh, S., Dialani, V., Mackey, L., Patterson, G.:
  Deepminer: Discovering interpretable representations for mammogram
  classification and explanation. arXiv preprint arXiv:1805.12323  (2018)

\bibitem{yilmaz2021automated}
Yilmaz, E.B., Buerger, C., Fricke, T., Sagar, M.M.R., Pe{\~n}a, J., Lorenz, C.,
  Gl{\"u}er, C.C., Meyer, C.: Automated deep learning-based detection of
  osteoporotic fractures in ct images. In: International Workshop on Machine
  Learning in Medical Imaging. pp. 376--385. Springer (2021)

\bibitem{yilmaz20203dunet_feat_attr}
Yilmaz, E.B., Mader, A.O., Fricke, T., Pe{\~n}a, J., Gl{\"u}er, C.C., Meyer,
  C.: Assessing attribution maps for explaining cnn-based vertebral fracture
  classifiers. In: Interpretable and Annotation-Efficient Learning for Medical
  Image Computing, pp. 3--12. Springer (2020)

\bibitem{zhang2021fine}
Zhang, Y., Khakzar, A., Li, Y., Farshad, A., Kim, S.T., Navab, N.: Fine-grained
  neural network explanation by identifying input features with predictive
  information. Advances in Neural Information Processing Systems  \textbf{34}
  (2021)

\bibitem{zhou2016learning}
Zhou, B., Khosla, A., Lapedriza, A., Oliva, A., Torralba, A.: Learning deep
  features for discriminative localization. In: Proceedings of the IEEE
  conference on computer vision and pattern recognition. pp. 2921--2929 (2016)

\end{thebibliography}

\end{document}